\documentclass[a4paper,11pt,fleqn]{article} 

\usepackage{amsmath,amssymb,amsthm,comment}

\allowdisplaybreaks

\makeatletter
\long\def\@makecaption#1#2{%
  \vskip\abovecaptionskip\footnotesize
  \sbox\@tempboxa{#1. #2}%
  \ifdim \wd\@tempboxa >\hsize
    #1. #2\par
  \else
    \global \@minipagefalse
    \hb@xt@\hsize{\hfil\box\@tempboxa\hfil}%
  \fi
  \vskip\belowcaptionskip}
\makeatother

\newcommand{\todo}[1][\null]{\ensuremath{\clubsuit}}

\newcommand{\noprint}[1]{}
\newcommand{\checked}[1][\null]{\ensuremath{\boldsymbol{\surd}}}

\newcommand{\p}{\partial}

\newcommand{\EqOrd}{r}

\newtheorem{theorem}{Theorem}

\newtheorem{corollary}[theorem]{Corollary}
\newtheorem{proposition}[theorem]{Proposition}
\newtheorem*{problem*}{Problem}
{\theoremstyle{definition}
\newtheorem{definition}[theorem]{Definition}

\newtheorem{remark}[theorem]{Remark}
\newtheorem*{remark*}{Remark}
}

\begin{document}

\par\noindent {\LARGE\bf
Equivalence groupoid of a class of \\general Burgers--Korteweg--de Vries equations\\
with space-dependent coefficients
\par}

{\vspace{4mm}\par\noindent {\large Stanislav Opanasenko
} \par\vspace{2mm}\par}

{\vspace{2mm}\par\noindent {\it
Department of Mathematics and Statistics, Memorial University of Newfoundland,\\
$\phantom{^{\dag}}$~St.\ John's (NL) A1C 5S7, Canada\par
}}
{\vspace{2mm}\par\noindent {\it
Institute of Mathematics of NAS of Ukra\"ine, 3 Tereshchenkivska Str., 01004 Ky\"iv, Ukra\"ine\par
}}

{\vspace{2mm}\par\noindent {\it
\textup{E-mail:} sopanasenko@mun.ca
}\par}

\vspace{8mm}\par\noindent\hspace*{10mm}\parbox{140mm}{\small
We describe the equivalence groupoid of the class of general Burgers - Korteweg - de Vries equations with space-dependent coefficients.
This class is shown to reduce by a family of equivalence transformations to a subclass whose usual equivalence group is four-dimensional.
Classified are admissible transformations of this subclass and singled out its subclasses admitting maximal nontrivial conditional equivalence groups.
All of them turn out to have dimension higher than four. In particular, a few new examples of nontrivial cases of normalization in the generalized 
sense of classes of differential equations appeared this way.
    }\par\vspace{4mm}

\noprint{
}

\section{Introduction}
A number of evolution equations that are important in mathematical physics are of the general form
\begin{equation}\label{Opanasenko:eq:GenBurgersKdVEqs}
 u_t+C(t,x)uu_x=\sum_{k=0}^\EqOrd A^k(t,x)u_k+B(t,x).
\end{equation}
In particular, this includes Burgers, Korteweg--de Vries (KdV), Kura\-mo\-to--Sivashinsky, Kawahara,
and generalized Burgers--KdV equations.

Here and in the following the integer parameter~$\EqOrd$ is fixed, and $\EqOrd\geqslant2$.
We require the condition $CA^\EqOrd\ne0$ guaranteeing
that equations from the class~\eqref{Opanasenko:eq:GenBurgersKdVEqs} are nonlinear and of genuine order~$\EqOrd$.
Throughout the paper we use the standard index derivative notation $u_t=\partial u/\partial t$, $u_k=\partial^k u/\partial x^k$.

The class~\eqref{Opanasenko:eq:GenBurgersKdVEqs} and its various subclasses were subject to studying from the symmetry analysis point of view,
see~\cite{Opanasenko:Opanasenko17} for an extensive list of references.
Recently, the class~\eqref{Opanasenko:eq:GenBurgersKdVEqs} became a source of examples of nontrivial equivalence groups~\cite{Opanasenko:Opanasenko17}.
In fact, the first examples of classes with generalized and extended generalized equivalence groups are of the form~\eqref{Opanasenko:eq:GenBurgersKdVEqs} (with some additional restrictions).
Moreover, detailed stu\-dying thereof allowed the authors to introduce the concept of an effective generalized
equivalence group of a class of differential equations.
Furthermore, the structure of this class is so flexible, that a ``reasonable'' singled
out subclass thereof is likely to possess normalization properties in some sense.
Nonetheless, it is not the case for a subclass~$\bar{\mathcal F}$ of equations with the
arbitrary elements being time-independent,
\begin{equation}\label{Opanasenko:eq:GenBurgersKdVEqsStat}
u_t+C(x)uu_x=\sum_{k=0}^\EqOrd A^k(x)u_k+B(x),\quad \text{where}\quad A^\EqOrd C\neq0.
\end{equation}
The aim of this paper is to thoroughly study admissible transforma\-tions of the class~$\bar{\mathcal F}$.
In a nutshell, the results of this paper comprise the following four facts.
Any equation in~$\bar{\mathcal F}$
is mapped  by an equivalence transformation of~$\bar{\mathcal F}$ to an equation in the subclass~$\mathcal F$ of reduced general Burgers--Korteweg--de Vries equations with
space-dependent coefficients, singled out by conditions $C=1$ and $A^1=0$.
The subclass~$\mathcal F$ is not normalized in any sense, and its usual equivalence group is four-dimen\-sional.
Classified are admissible transformations of the class~$\mathcal F$ and sing\-led out are its subclasses admitting maximal
nontrivial conditional equivalence subgroups of the equivalence group of~$\mathcal F$,
\begin{gather*}
\hat{\mathcal F}_{\mathrm I,1}\colon\quad u_t+uu_x=\left(\frac{\alpha+2}{a_{01}}b_1+a_{01}|x+\beta|^{\alpha}\right)u+\sum_{j=2}^\EqOrd a_j(x+\beta)^j|x+\beta|^{\alpha}u_j\\
\qquad\quad\ {}+(x+\beta)\left(b_2|x+\beta|^{2\alpha}+b_1|x+\beta|^{\alpha}-\frac{b_1^2(\alpha+1)}{a_{01}^2}\right)\quad\text{with } \quad\alpha a_ra_{01}\neq0,\\
\hat{\mathcal F}_{\mathrm I,01}\colon\quad u_t+uu_x=\sum_{j=2}^\EqOrd a_j(x+\beta)^j|x+\beta|^{\alpha}u_j+a_{00}u\\
\qquad\quad\ {}+(x+\beta)\left(b_2|x+\beta|^{2\alpha}-\frac{\alpha+1}{(\alpha+2)^2}a_{00}^2\right)\quad\text{with }\quad (\alpha+2) a_r\neq0,\\
\hat{\mathcal F}_{\mathrm I,00}\colon\quad u_t+uu_x=\sum_{j=2}^\EqOrd a_j(x+\beta)^{j-2}u_j+b_0(x+\beta)
+b_2(x+\beta)^{-5}\quad\text{with }\quad a_r\neq0,\\
\hat{\mathcal F}_{\mathrm {II},0}\colon\quad u_t+uu_x=\sum\limits_{j=2}^r a_j(x+\beta)^ju_j+a_{00}u+b_0\quad \text{with }\ a_r\neq0,\\
\hat{\mathcal F}_{\mathrm {II},1}\colon\quad u_t+uu_x=\sum\limits_{j=2}^r a_j(x+\beta)^ju_j+(a_{01}\ln |x+\beta|+a_{00})u\\
\qquad\quad\ {}+(x+\beta)\left(-\frac{a_{01}^2}4\ln^2 |x+\beta|+\left(\frac{a_{01}^2}4-\frac{a_{00}a_{01}}2\right)
\ln |x+\beta|+b_0\right)\ \ \text{with}\ \ a_ra_{01}\neq0,\\
\mathcal F_{\mathrm {III}}\colon\quad u_t+uu_x=\sum\limits_{j=2}^r a_je^{\alpha x}u_j+(a_{01}e^{\alpha x}+a_{00})u+b_2e^{2\alpha x}-\frac{a_{00}a_{01}}{\alpha}e^{\alpha x}-\frac{a_{00}^2+a_{00}}{2\alpha}\\ \qquad\quad\ {} \text{with } \ \alpha a_r\neq0,\\
\mathcal F_{\mathrm {IV},1}\colon\quad u_t+uu_x=\sum_{j=2}^r a_ju_j+a_0u+b_1x+b_0\quad\text{with }\quad \alpha a_r\sum_{j=2}^{r-1}|a_j|\neq0,\\
\mathcal F_{\mathrm {IV},0}^{\EqOrd>2}\colon\quad u_t+uu_x=a_ru_r+a_0u+\frac{r-1}{(r-2)^2}a_0^2 x+b_0\quad\text{with } \quad\alpha a_r\neq0,\ r>2,\\
\mathcal F_{\mathrm {IV},0}^{\EqOrd=2}\colon\quad u_t+uu_x=a_2u_2+b_1x+b_0\quad\text{with }\quad \alpha a_r\neq0.
\end{gather*}
All these subclasses but~$\hat{\mathcal F}_{\mathrm{II},0}$ are normalized in the generalized sense.
The class~$\hat{\mathcal F}_{\mathrm{II},0}$ is normalized in the usual sense.

The main result of the paper is described in the following theorem.

\begin{theorem}
The usual equivalence group of the class~$\mathcal F$ of reduced general Burgers--Korteweg--de Vries equations with
space-dependent co\-efficients is four-dimensional. The list of maximal nontrivial conditional equivalence subgroups is exhausted by the
generalized equivalence groups of the normalized subclasses $\hat{\mathcal F}_{\mathrm I,1}$, $\hat{\mathcal F}_{\mathrm I,01}$, $\hat{\mathcal F}_{\mathrm I,00}$, $\hat{\mathcal F}_{\mathrm {II},1}$, $\mathcal F_{\mathrm {III}}$, $\mathcal F_{\mathrm {IV},1}$, $\mathcal F_{\mathrm {IV},0}^{\EqOrd>2}$, $\mathcal F_{\mathrm {IV},0}^{\EqOrd=2}$ and the usual equivalence group of the normalized subclass~$\hat{\mathcal F}_{\mathrm {II},0}$. The equivalence groupoid of the class~$\mathcal F$ is generated by its usual equivalence group and the equivalence groups of the above subclasses.
\end{theorem}

For all classes normalized in the generalized sense, we can take their effective generalized equivalence subgroups as
maximal conditional equivalence groups.
Denote by~$\mathcal F_0$ the complement to the union of the above subclasses in the class~$\mathcal F$. It is a normalized class in the usual sense, and its equivalence group coincides with that of~$\mathcal F$.

\begin{corollary}
The class~$\mathcal F$ is a union of the normalized $($in either the generalized or the usual sense$)$ classes~$\hat{\mathcal F}_{\mathrm I,1}$,
$\hat{\mathcal F}_{\mathrm I,01}$, $\hat{\mathcal F}_{\mathrm I,00}$, $\hat{\mathcal F}_{\mathrm {II},1}$, $\hat{\mathcal F}_{\mathrm {II},0}$, $\mathcal F_{\mathrm {III}}$, $\mathcal F_{\mathrm {IV},1}$, $\mathcal F_{\mathrm {IV},0}^{\EqOrd>2}$, $\mathcal F_{\mathrm {IV},0}^{\EqOrd=2}$
and~$\mathcal F_0$.
\end{corollary}

The structure of this paper is as follows. Firstly, we remind in Sec\-tion~2
theoretical foundations related to an equivalence within classes of differential equations. Following~\cite{Opanasenko:Opanasenko17} in Section~3 we recall the structure of the equivalence
groupoids of the superclass of general Burgers--Kor\-te\-weg--de Vries equations, its subclass of equations with time-independent
coefficients and the gauging of these classes to the corresponding sub\-classes of reduced equations. In Section~4 we give the complete classification of admissible transformations of the class~$\mathcal F$ of reduced general Burgers--KdV equations with space-dependent coefficients. In~\cite{Opanasenko:Opanasenko17} there were found subclasses of the class~$\mathcal F$
admitting admissible transformations that are not generated by the equivalence transformations of~$\mathcal F$. But the question of a structure of equivalence groupoids
of these subgroups was not addressed there. Here we fill this gap by comprehensively describing all these subclasses and
their equivalence groups (for subclasses normalized in the generalized sense we present either the entire generalized equivalence
group, or its effective generalized equivalence group or both of them). By partitioning if necessary these subclasses we achieve a normalization of ``subsubclasses'' in either usual or generalized sense.
Thus we present the superclass~$\mathcal F$ as a union of normalized classes of differential equations described in Theorem~1.
For the two normalized subclasses to be able to have a closed form of group transformations we apply a non-standard approach,
the technical crux of which is as follows. First we gauge the class under consideration by a family of equivalence transformations thereof to a
nice normalized subclass. Then every equivalence transformation in the class under consideration would be a composition of
the gauging mapping, an equivalence transformation within the nice subclass and the inverse of a (not the same as before as
we consider not symmetry but equivalence transformations of the superclass) gauging mapping.
This procedure may explain an appearance of generalized equivalence groups for most of the considered subclasses.
In fact, the determining systems of ODEs are exactly solvable for all but the two equivalence groups and this procedure is only
lurking in the background, but we could use it almost everywhere. In this case, even if a nice underlying subclass is normalized in
the usual sense, we compose its equivalence transformations with transformations from the families parameterized by arbitrary elements of the
superclass, and thus parameterize the equivalence transformations thereof by arbitrary elements, making them generalized ones.

\section{Equivalence of classes of differential equations}
We recall the essential notions for the present paper only. See~\cite{Opanasenko:Opanasenko17,Opanasenko:popo06Ay,Opanasenko:popo10Ay} for more details.
Let $\mathcal L_\theta$ denote a system of differential equations of the form
\[L\big(x,u^{(\EqOrd)},\theta\big(x,u^{(\EqOrd)}\big)\big)=0,\]
where  $x=(x_1,\dots,x_n)$ is the~$n$ independent variables, $u=(u^1,\dots,u^m)$ is the~$m$ dependent variables, and~$L$ is a tuple of differential functions in~$u$.
We use the standard short-hand notation~$u^{(\EqOrd)}$ to denote the tuple of derivatives of~$u$ with respect to $x$ up to order~$\EqOrd$,
which also includes the~$u$'s as the derivatives of order zero.
The system~$\mathcal L_\theta$ is parameterized by the tuple of functions~$\theta=(\theta^1(x,u^{(\EqOrd)}),\dots,\theta^k(x,u^{(\EqOrd)}))$, called the arbitrary elements running through the solution set~$\mathcal S$ of an auxiliary system
of differential relations in~$\theta$.
Thus, the \textit{class of $($systems of$)$ differential equa\-tions} $\mathcal L|_{\mathcal S}$ is the parameterized family of systems~$\mathcal L_\theta$'s, such that~$\theta$ lies in~$\mathcal S$.

\looseness=-1 Equivalence of classes of differential equations is based on studying how equations from a given class are mapped to each other.
The notion of \textit{admissible transformations},
which constitute the \textit{equivalence groupoid} of the class~$\mathcal L|_{\mathcal S}$,
formalizes this study.
An admissible transformation is a triple $(\theta,\tilde\theta,\varphi)$,
where $\theta,\tilde\theta\in\mathcal S$ are arbitrary-element tuples
associated with equations $\mathcal L_\theta$ and~$\mathcal L_{\tilde\theta}$ from the class~$\mathcal L_{\mathcal S}$
that are similar to each other,
and $\varphi$ is a point transformation in the space of~$(x,u)$ that maps~$\mathcal L_\theta$ to~$\mathcal L_{\tilde \theta}$.

A related notion of relevance in the group classification of differential equations is that of \textit{equivalence transformations}.
Usual equivalence transformations are point transformations
in the joint space of indepen\-dent variables, derivatives of~$u$ up to order~$\EqOrd$ and arbitrary elements
that are projectable to the space of~$(x,u^{(\EqOrd')})$ for each $\EqOrd'=0,\dots,\EqOrd$,
respect the contact structure of the $\EqOrd$th order jet space coordinatized by the $\EqOrd$-jets $(x,u^{(\EqOrd)})$
and map every system from the class~$\mathcal L|_{\mathcal S}$ to a system from the same class.
The Lie (pseudo)group constituted by the equivalence transformations of~$\mathcal L|_{\mathcal S}$
is called the \emph{usual equivalence group} of this class and denoted by~$G^\sim$.

Each equivalence transformation~$\mathcal T\in G^\sim$ generates
a family of admis\-sible transformations parameterized by~$\theta$,
\[
G^\sim\ni\mathcal T\rightarrow\big\{(\theta,\mathcal T\theta,\pi_*\mathcal T)\,|\, \theta\in\mathcal S\big\}\subset\mathcal G^\sim,
\]
and therefore the usual equivalence group~$G^\sim$ gives rise to a subgroupoid of the equivalence groupoid~$\mathcal G^\sim$.
The function~$\pi$ is the projection of the space of $(x,u^{(\EqOrd)},\theta)$ to the space of equation variables only, $\pi(x,u^{(\EqOrd)},\theta)=(x,u)$.
The pushforward $\pi_*\mathcal T$ of~$\mathcal T$ by $\pi$ is then just the restriction of~$\mathcal T$ to the space of~$(x,u)$.

The projectability property for equivalence transformations can be neglected.
Then these equivalence transformations constitute a Lie pse\-u\-do\-group~$\bar G^{\sim}$
called the \emph{generalized equivalence group} of the class.
See the first discussion of this notion in~\cite{Opanasenko:mele94Ay,Opanasenko:mele96Ay} and the further development in~\cite{Opanasenko:popo06Ay,Opanasenko:popo10Ay}.
When the generalized equivalence group coincides with the usual one
the situation is considered to be trivial.
Similar to usual equivalence transformations,
each element of~$\bar G^{\sim}$ generates a family of admissible transformations parameterized by~$\theta$,
\[
\bar G^\sim\ni\mathcal T\rightarrow\big\{(\theta',\mathcal T\theta',
\pi_*(\mathcal T|_{\theta=\theta'(x,u)}))\,|\, \theta'\in\mathcal S\big\}\subset\mathcal G^\sim,
\]
and thus the generalized equivalence group~$\bar G^{\sim}$ also generates
a sub\-grou\-poid~$\bar{\mathcal H}$ of the equivalence groupoid~$\mathcal G^\sim$.

\begin{definition}
Any minimal subgroup of~$\bar G^{\sim}$
that generates the same subgroupoid of~$\mathcal G^\sim$
as the entire group~$\bar G^{\sim}$ does
is called an \emph{effective generalized equivalence group} of the class~$\mathcal L|_{\mathcal S}$.
\end{definition}

If the entire group~$\bar G^{\sim}$ is effective itself, then its uniqueness is evident.
At the same time, there exist classes of differential equations,
where effective generalized equivalence groups are proper subgroups
of the corresponding generalized equivalence groups that are even not normal.
Hence each of these effective generalized equivalence groups is not unique
since it differs from some of subgroups non-identically similar to it,
and all of these subgroups are also effective generalized equivalence groups of the same class.

The class of differential equations~$\mathcal L|_{\mathcal S}$ is \emph{normalized}
in the usual (resp.\ generalized) sense
if the subgroupoid induced by its usual (resp.\ generalized) equivalence group
coincides with the entire equivalence groupoid~$\mathcal G^\sim$ of~$\mathcal L|_{\mathcal S}$.
The normalization of~$\mathcal L|_{\mathcal S}$ in the usual sense
is equivalent to the following conditions.
The transformational part~$\varphi$ of each admissible transformation $(\theta',\theta'',\varphi)\in\mathcal G^\sim$
does not depend on the fixed initial value~$\theta'$ of the arbitrary-element tuple~$\theta$
and, therefore, is appropriate for any initial value of~$\theta$.

The normalization properties of the class~$\mathcal L|_{\mathcal S}$
are usually established via computing its equivalence groupoid~$\mathcal G^\sim$,
which is realized using the direct method.
Here one fixes two arbitrary systems from the class,
$\mathcal L_\theta\colon L(x,u^{(\EqOrd)},\theta(x,u^{(\EqOrd)}))=0$
and $\mathcal L_{\tilde\theta}\colon L(\tilde x,\tilde u^{(\EqOrd)},\tilde \theta(\tilde x,\tilde u^{(\EqOrd)}))=0$,
and aims to find the (nondegenerate) point transformations,
$\varphi$: $\tilde x_i=X^i(x,u)$, $\tilde u^a=U^a(x,u)$, $i=1,\dots,n$, $a=1,\dots,m$, connecting them.
For this, one changes the variables in the system~$\mathcal L_{\tilde\theta}$
by expressing the derivatives $\tilde u^{(\EqOrd)}$ in terms of $u^{(\EqOrd)}$ and derivatives of the functions $X^i$ and $U^a$
as well as by substituting $X^i$ and $U^a$ for $\tilde x_i$ and $\tilde u^a$, respectively.
The requirement that the resulting transformed system
has to be satisfied identically for solutions of~$\mathcal L_\theta$
leads to the system of determining equations for the components of the transformation~$\varphi$.

Imposing additional constraints on arbitrary elements of the class, we may single out its
subclass whose equivalence group is not contained in the equivalence
group of the entire class. Let $\mathcal L|_{\mathcal S'}$ be the subclass of the class $\mathcal L|_{\mathcal S}$, which is constrained
by the additional system of equations $\mathcal S'(x, u^{(r)}, \theta^{(q')}) = 0$ and inequalities $\Sigma'(x, u^{(r)}, \theta^{(q')}) \neq 0$
with respect to the arbitrary elements $\theta = \theta(x, u^{(r)})$. Here $\mathcal S' \subset \mathcal S$ is
the set of solutions of the united system $\mathcal S = 0$, $\Sigma \neq0$, $\mathcal S' = 0$, $\Sigma' \neq 0$. We assume that the
united system is compatible for the subclass $\mathcal L|_{\mathcal S'}$ to be nonempty.

\begin{definition}
The equivalence group $G^\sim(\mathcal L|_{\mathcal S'})$ of the subclass $\mathcal L|_{\mathcal S'}$ is called a conditional equivalence
group of the entire class $\mathcal L|_{\mathcal S}$ under the conditions $\mathcal S' = 0$, $\Sigma' \neq 0$. The conditional
equivalence group is called nontrivial if it is not a subgroup of $G^\sim(\mathcal L|_{\mathcal S})$.
\end{definition}

Conditional equivalence groups may be trivial not with respect to the equivalence group of
the entire class but with respect to other conditional equivalence groups. Indeed, if $\mathcal S' \subset \mathcal S''$ and
$G^\sim(\mathcal L|_{\mathcal S'})\subset G^\sim(\mathcal L|_{\mathcal S''})$ then the subclass $\mathcal L|_{\mathcal S'}$ is not interesting
from the conditional symmetry point of view. Therefore, the set of additional conditions on the arbitrary elements can be
reduced substantially.

\begin{definition}
 The conditional equivalence group $G^\sim_{\mathcal L|_{\mathcal S'}}$ of the class~$\mathcal L|_{\mathcal S}$ under the additional
conditions $\mathcal S' = 0$, $\Sigma'\neq 0$ is called maximal if for any subclass $\mathcal L|_{\mathcal S''}$ of the class $\mathcal L|_{\mathcal S}$ containing
the subclass $\mathcal L|_{\mathcal S'}$ we have $G^\sim_{\mathcal L|_{\mathcal S'}} \not\subset G^\sim_{\mathcal L|_{\mathcal S''}}$.
\end{definition}

\section{Preliminary analysis of equivalence groupoid}
We start studying admissible transformations of the class~$\mathcal F$ by presenting the equivalence groupoid of its
superclass~\eqref{Opanasenko:eq:GenBurgersKdVEqs} and then descend there\-from to the class under study.

\begin{proposition}\label{Opanasenko:pro:EquivGroupoidOfGenBurgersKdVEqs}
The class~\eqref{Opanasenko:eq:GenBurgersKdVEqs} is normalized in the usual sense.
Its usual equivalence group~$G^\sim_{\mbox{\tiny\eqref{Opanasenko:eq:GenBurgersKdVEqs}}}$
consists of the transformations in the joint space of $(t,x,u,\theta)$
whose $(t,x,u)$-components are of the form
\[
\tilde t=T(t),\quad \tilde x=X(t,x),\quad \tilde u=U^1(t)u+U^0(t,x),
\]
where $T=T(t)$, $X=X(t,x)$, $U^1=U^1(t)$ and $U^0=U^0(t,x)$
are arbitrary smooth functions of their arguments such that $T_tX_xU^1\ne0$.
\end{proposition}

Following~\cite{Opanasenko:Opanasenko17} we can gauge the arbitrary elements~$C=1$ and~$A^1=0$ by a family of equivalence
transformations of the class~\eqref{Opanasenko:eq:GenBurgersKdVEqs} and obtain the class of reduced general Burgers--KdV equations
\begin{gather}\label{Opanasenko:eq:GenBurgersKdVEqsGaugedSubclass}
u_t+uu_x=\sum_{j=2}^\EqOrd A^j(t,x)u_j+A^0(t,x)u+B(t,x).
\end{gather}
As before, the arbitrary elements run through the set of smooth functions of~$(t,x)$ with $A^\EqOrd C\neq0$.

\begin{theorem}\label{Opanasenko:thm:EquivalenceGroupGenBurgersKdVEqsGaugeC1A10}
The class
of reduced $(1{+}1)$-dimensional general $\EqOrd$th~order Burgers--KdV equations~\eqref{Opanasenko:eq:GenBurgersKdVEqsGaugedSubclass}
is normalized in the usual sense.
Its usual equivalence group~$G^\sim$ consists of the transformations of the form
\begin{gather}
\tilde t=T(t),\quad \tilde x=X^1(t)x+X^0(t),\quad \tilde u=\frac{X^1}{T_t}u+\frac{X^1_t}{T_t}x+\frac{X^0_t}{T_t},
\label{Opanasenko:eq:PointTransformationBetweenGenBurgersKdVEqsGaugeC1A10a}\\
\tilde A^j=\frac{(X^1)^j}{T_t}A^j,\quad  \tilde A^0=\frac{1}{T_t}\left(A^0+2\frac{X^1_t}{X^1}-\frac{T_{tt}}{T_t}\right),
\label{Opanasenko:eq:PointTransformationBetweenGenBurgersKdVEqsGaugeC1A10b}\\
\tilde B=\frac{X^1}{(T_t)^2}B+\frac1{T_t}\left(\frac{X^1_t}{T_t}\right)_tx+\frac1{T_t}\left(\frac{X^0_t}{T_t}\right)_t
-\left(\frac{X^1_t}{T_t}x+\frac{X^0_t}{T_t}\right)\tilde A^0,\!\!\!
\label{Opanasenko:eq:PointTransformationBetweenGenBurgersKdVEqsGaugeC1A10c}
\end{gather}
where $j=2,\dots,\EqOrd$, and $T=T(t)$, $X^1=X^1(t)$ and $X^0=X^0(t)$ are arbitrary smooth functions of their arguments with $T_tX^1\ne0$.
\end{theorem}

The subclass~$\bar{\mathcal F}$ of general Burgers--KdV equations with space-depen\-dent coefficients
is singled out from the class~\eqref{Opanasenko:eq:GenBurgersKdVEqs}
by the constraints $A^k_t=0$, $k=0,\dots,\EqOrd$, $B_t=0$ and $C_t=0$.
Therefore, its usual equivalence group~$G^\sim_{\bar{\mathcal F}}$
is a subgroup of~$G^\sim_{\mbox{\tiny\eqref{Opanasenko:eq:GenBurgersKdVEqs}}}$
that consists of transformations preserving the above constraints.

\begin{proposition}\label{Opanasenko:pro:EquivGroupOfGenBurgersKdVEqsWithX-dependentCoeffs}
The usual equivalence group~$G^\sim_{\bar{\mathcal F}}$ of
the class~$\bar{\mathcal F}$ of general Burgers--Korteweg--de Vries equations
with space-dependent co\-efficients
consists of the transformations in the joint space of $(t,x,u,\theta)$
whose $(t,x,u)$-components are of the form
\[
\tilde t=c_1t+c_2,\quad \tilde x=X(x),\quad \tilde u=c_3'u+U^0(x),
\]
where $c_1$, $c_2$ and~$c_3'$ are arbitrary constants
and $X=X(x)$ and~$U^0=U^0(x)$ are arbitrary smooth functions of~$x$ such that $c_1X_xc_3'\ne0$.
\end{proposition}

The existence of classifying conditions~\cite{Opanasenko:Opanasenko17}
\[
\frac{T_t}{(X_x)^\EqOrd}X_t\tilde A^\EqOrd_{\tilde x}+\left(\frac{T_t}{(X_x)^\EqOrd}\right)_t\tilde A^\EqOrd=0,\quad
\frac{T_tU^1}{X_x}X_t\tilde C_{\tilde x}+\left(\frac{T_tU^1}{X_x}\right)_t\tilde C=0,
\]
for admissible transformations of the class~$\bar{\mathcal F}$ implies that it is definitely not normalized in any sense.
At the same time, we can once again gauge the arbitrary elements $C$ and~$A^1$
by means of equivalence transformations of the class~$\bar{\mathcal F}$ and produce the class~$\mathcal F$
of reduced general Burgers--KdV equations with space-dependent coefficients
\[u_t+uu_x=\sum_{j=2}^\EqOrd A^j(x)u_j+A^0(x)u+B(x).\]

\begin{proposition}
The usual equivalence group~$G^\sim_{\mathcal F}$ of the class~$\mathcal F$ is four-dimensional
and consists of the transformations of the form
\begin{gather*}
\tilde t=c_1t+c_2,\quad
\tilde x=c_3x+c_4,\quad
\tilde u=\frac{c_3}{c_1}u,\quad
\tilde A^j=\frac{c_3^{j}}{c_1}A^j,\quad
\tilde A^0=\frac1{c_1}A^0,\quad
\tilde B=\frac{c_3}{c_1^2}B,
\end{gather*}
where $j=2,\dots,\EqOrd$,
and $c$'s are arbitrary constants with $c_1c_3\ne0$.
\end{proposition}

Neither the class~$\bar{\mathcal F}$ nor its superclass~$\bar{\mathcal F}$ are normalized in any sense.
Thus, the problem of describing the equivalence groupoid~$\mathcal G^\sim_{\mathcal F}$ of the class~$\mathcal F$
should be considered as the classification of admissible trans\-forma\-tions
up to~$G^\sim_{\mathcal F}$-equivalence, see~\cite[Sections~2.6 and~3.4]{Opanasenko:popo10Ay}.
The class~$\mathcal F$ is a subclass of the class~\eqref{Opanasenko:eq:GenBurgersKdVEqsGaugedSubclass},
whence~$\mathcal G^\sim_{\mathcal F}$
is a subgroupoid of the equivalence groupoid of the class~\eqref{Opanasenko:eq:GenBurgersKdVEqsGaugedSubclass},
and the results of Theorem~\ref{Opanasenko:thm:EquivalenceGroupGenBurgersKdVEqsGaugeC1A10}
are valid here, although they should be further specified.
This is achieved by differentiating the relations~\eqref{Opanasenko:eq:PointTransformationBetweenGenBurgersKdVEqsGaugeC1A10b}--\eqref{Opanasenko:eq:PointTransformationBetweenGenBurgersKdVEqsGaugeC1A10c},
solved with respect to the source arbitrary elements, with respect to~$t$.
This gives the classifying conditions for admissible transformations,
\begin{gather}
\big(X^1_tx+X^0_t\big)\tilde A^j_{\tilde x}+\left(\frac{T_{tt}}{T_t}-j\frac{X^1_t}{X^1}\right)\tilde A^j=0,
\label{Opanasenko:eq:GenBurgersKdVEqsWithX-dependentCoeffsClassifyingEqsForAdmTransA}\\
\big(X^1_tx+X^0_t\big)\tilde A^0_{\tilde x}+\frac{T_{tt}}{T_t}\tilde A^0=
\frac1{T_t}\left(2\frac{X^1_t}{X^1}-\frac{T_{tt}}{T_t}\right)_t,\
\label{Opanasenko:eq:GenBurgersKdVEqsWithX-dependentCoeffsClassifyingEqsForAdmTransB}\\
\big(X^1_tx+X^0_t\big)\tilde B_{\tilde x}+\left(2\frac{T_{tt}}{T_t}-\frac{X^1_t}{X^1}\right)\tilde B=
-\frac{T_t}{X^1}\big(X^1_tx+X^0_t\big)^2\tilde A^0_{\tilde x}\nonumber\\
\quad{}
-\frac{X^1}{T_t^2}\left(T_t\frac{X^1_tx+X^0_t}{X^1}\right)_t\tilde A^0
+\frac{X^1}{T_t^2}\left(\frac{T_t}{X^1}\left(\frac{X^1_tx+X^0_t}{T_t}\right)_t\right)_t,\label{Opanasenko:eq:GenBurgersKdVEqsWithX-dependentCoeffsClassifyingEqsForAdmTransC}
\end{gather}
where the initial space variable~$x$ should be substituted, after expanding all derivatives,
by its expression via~$\tilde x$, $x=(\tilde x-X^0)/X^1$.
Note that admissible transformations with $T_{tt}=X^0_t=X^1_t=0$ are generated
by the usual equivalence group~$G^\sim_{\mathcal F}$.

\section{Nontrivial conditional equivalence subgroups}
In~\cite{Opanasenko:Opanasenko17} with a~help of the method of furcate splitting, cf.~\cite{Opanasenko:NikPop,Opanasenko:Opanasenko19}, solved were the
classifying conditions~\eqref{Opanasenko:eq:GenBurgersKdVEqsWithX-dependentCoeffsClassifyingEqsForAdmTransA}--\eqref{Opanasenko:eq:GenBurgersKdVEqsWithX-dependentCoeffsClassifyingEqsForAdmTransC}
for admissible transformations of the class~$\mathcal F$, but the obtained
admissible transformations were presented superficially. More precisely, they were parameterized by solutions of some ODEs.
Here we study the question in more depth and present explicit forms of group parameters of the nontrivial conditional
equivalence groups.
Besides, following~\cite{Opanasenko:Opanasenko17} for simplicity we consider only subclasses of the classes~$\mathcal F_{\rm I}$ and~$\mathcal F_{\rm {II}}$,
defined below, admitting proper subgroups of maximal conditional equivalence groups. In fact, these subgroups are the quotients thereof
by the space-translations. Note that given in Theorem~1 are the subclasses admitting maximal nontrivial conditional equivalence
subgroups.

\medskip

\noindent$\mathbf {I.}$ The class~$\mathcal F_{\mathrm I}$ of equations
\begin{gather*}
u_t+uu_x=\sum_{j=2}^\EqOrd a_jx^j|x|^{\alpha}u_j+\big(a_{00}+a_{01}|x|^{\alpha}\big)u+x\big(b_0+b_1|x|^{\alpha}+b_2|x|^{2\alpha}\big)
\end{gather*}
with $\alpha a_r\neq0$ naturally partitions into the two $\mathcal G^\sim_{\mathcal F_{\mathrm I}}$-invariant subclasses $\mathcal F_{\mathrm {I},0}$ and
$\mathcal F_{\mathrm{I},1}$ singled out by the conditions $a_{01}=0$ and $a_{01}\neq0$, respectively, since the arbitrary
element~$a_{01}$ is easily shown to be transformed by the rule $\tilde a_{01}=c_4a_{01}$ under admissible transformations of the
class, $c_4\neq0$. The class $\mathcal F_{\mathrm I,1}$ admits additional admissible transformations if and only if
$a_{00}=(\alpha+2)b_1/a_{01}$ and $b_0=-b_1^2(1+\alpha)/a_{01}^2$, so we reduce the arbitrary-elements tuple of the class
by $a_{00}$ and $b_0$ and denote the subclass obtained again by~$\mathcal F_{\mathrm I,1}$.

\begin{proposition}~\label{Opanasenko:prop:BKdVStF_1}
The class~$\mathcal F_{\mathrm I,1}$ is normalized in the generalized sense. Its generalized equivalence group
consists of the point transformations in the relevant space, which are of the form
\begin{gather*}
\tilde t=\bar T,\quad \tilde x=\bar X^1 x,\quad \tilde u=\frac{\bar X^1}{\bar T_t}u-\frac{\bar X^1_{t}}{\bar T_t}x,\\
\tilde \alpha=\alpha,\quad\tilde a_j=\bar c_4 a_j,\quad \tilde a_{01}=\bar c_4 a_{01},\quad
\tilde b_2=\bar c_4^2 b_2,\quad \tilde b_1=\bar c_5,
\end{gather*}
where $\bar T$ is a smooth function of~$t$ and the arbitrary elements~$\theta=(\alpha,a_j,a_{01},b_2,b_1)$,
\begin{gather*}
\bar T(t,\theta)=\frac{1}{\bar c_5}\ln\left|\bar c_5
\left(c_1\frac{e^{-b_1\alpha t/a_{01}}-1}{-b_1\alpha/a_{01}}+c_2\right)+1\right|,
\end{gather*}
taking the form at the singular points
\begin{gather*}
\begin{array}{ll}
\bar T(t,\theta)=\bar c_1\dfrac{e^{-b_1\alpha t/a_{01}}-1}{-b_1\alpha/a_{01}}+\bar c_2  &\text{if }\quad \bar c_5=0 \text{ and } b_1\neq0,\\[2ex]
\bar T(t,\theta)=\dfrac1{\bar c_5}\ln|{\bar c_5}\left(\bar c_1t+\bar c_2\right)|        &\text{if }\quad\bar c_5\neq0 \text{ and } b_1=0,\\[2ex]
\bar T(t,\theta)=\bar c_1{t}+\bar c_2                                                  &\text{if }\quad(\bar c_5,b_1)=(0,0),
\end{array}
\end{gather*}
$\bar c$'s are arbitrary functions of~$\theta$ with
$\bar c_1\bar c_4\frac{\p(\tilde a_2,\dots,\tilde a_r,\tilde a_{01},\tilde b_1,\tilde b_2)}{\p(a_2,\dots,a_r,a_{01},b_1,b_2)}\neq0$ as well as
$\bar X^1(t,\theta)=(\bar c_4\bar T_t)^{-1/\alpha}$ if $\alpha$ is odd or rational in the reduced form with an odd numerator
and $\bar X^1(t)=\varepsilon|\bar c_4\bar T_t|^{-1/\alpha}$ with $\varepsilon=\pm1$ and $\bar c_4\bar T_t>0$ otherwise.
\end{proposition}

\begin{remark}
The function~$T$ is a solution of an ODE smoothly depending on parameters, so it is a smooth function of these parameters
and initial conditions~\cite[Corollary~6, p.~97]{Opanasenko:Arnold1992}
($\alpha$, $b_1$ and~$a_{01}$ are the parameters of the equation in this case, $c$'s are the initial conditions).
This argumentation is valid for the group parameters in the equivalence groups below, where appropriate, as well.
In fact, in these cases it follows from the transformation for~$\tilde A^0$ (the equation~\eqref{Opanasenko:eq:PointTransformationBetweenGenBurgersKdVEqsGaugeC1A10b}) that the function~$T$
satisfies the equation
\[
\gamma=\delta\frac1{T_t}+\left(\frac1{T_t}\right)_t=0
\]
for some constants~$\gamma$ and~$\delta$, having the general solution
\[
T(t)=\frac1\gamma\ln\left|\gamma\left(c_1\frac{e^{\delta t}-1}{\delta}+c_2\right)+1\right|.
\]
The continuity of this function is evident and at the singular points the function takes the form
\begin{gather*}
\begin{array}{ll}
T(t)=c_1\dfrac{e^{\delta t}-1}{\delta}+c_2\ &\text{ if }\quad \gamma=0\text{ and } \delta\neq0,\\[2ex]
T(t)=\dfrac1\gamma\ln|\gamma(c_1t+c_2)+1|  &\text{ if }\quad \gamma\neq0\text{ and } \delta=0,\\[2ex]
T(t)=c_1t+c_2&\text{ if }\quad (\gamma,\delta)=(0,0).
\end{array}
\end{gather*}
The transformations in Proposition~\ref{Opanasenko:prop:BKdVStF_1} indeed form a group, which is straightforward to show.
Therefore the equivalence group of the class~$\mathcal F_{\mathrm I,1}$ is a local Lie group of transformations
(all equivalence group here and below in the paper are finite-dimensional so we do not need to talk about Lie pseudogroups).
If the function~$T$ is of the form~$\frac1\gamma\ln|\gamma(c_1t+c_2)+1|$ and $\gamma c_2=-1$, then $T(t)$ degenerates into an affine function. To avoid this, in all such situations thereafter we implicitly assume otherwise.
The notation $\frac{\p(\cdot,\dots,\cdot)}{\p(\cdot,\dots,\cdot)}$ stands for the determinant of the corresponding Jacobian matrix. Thereafter, we will not call attention to these facts.
\end{remark}

Since the arbitrary element~$\alpha$ is invariant under admissible trans\-forma\-tions, it is convenient to consider the two
subclasses~$\mathcal F_{\mathrm {I},00}$ and~$\mathcal F_{\mathrm {I},01}$ of~$\mathcal F_{\mathrm {I},0}$
singled out by conditions $\alpha=-2$ and $\alpha\neq-2$, respectively. To achieve an extension of a number of admissible
transformations in the latter class we need to consider its subclass (denoted again~$\mathcal F_{\mathrm {I},01}$) singled
out by the conditions $b_1=0$ and $b_0=-(\alpha+1)a_{00}^2/(\alpha+2)^2$.

\begin{proposition}
The class~$\mathcal F_{\mathrm I,01}$ is normalized in the generalized sense. Its generalized equivalence group
consists of the point transformations of the form
\begin{gather*}
\tilde t=\bar T(t),\quad \tilde x=\bar X^1(t) x,\quad \tilde u=\frac{\bar X^1}{\bar T_t}u-\frac{\bar X^1_{t}}{\bar T_t}x,\\
\tilde \alpha=\alpha,\quad\tilde a_j=\bar c_4 a_j,\quad \tilde a_{00}=\bar c_5,\quad
\tilde b_2=\bar c_4^2 b_2,
\end{gather*}
where $\bar T$ is a smooth function of~$t$ and the arbitrary elements~$\theta$,
\[
\bar T(t,\theta)=\frac1{\bar c_5}\ln\left|\bar c_5\left(\bar c_1\frac{e^{a_{00}\alpha t/(\alpha+2)}-1}{a_{00}\alpha/(\alpha+2)}+\bar c_2\right)+1\right|.
\]
The function~$\bar T$ takes at the singular points the following forms
\begin{gather*}
\begin{array}{ll}
\bar T(t,\theta)=\bar c_1\dfrac{e^{a_{00}\alpha t/(\alpha+2)}-1}{a_{00}\alpha/(\alpha+2)}+\bar c_2 &\text{ if }\quad \bar c_5=0\text{ and } a_{00}\neq0,\\[2ex]
\bar T(t,\theta)=\dfrac1{\bar c_5}\ln|\bar c_5(\bar c_1t+\bar c_2)+1|&\text{ if }\quad \bar c_5\neq0\text{ and } a_{00}=0,\\[2ex]
\bar T(t,\theta)=\bar c_1t+\bar c_2&\text{ if }\quad (\bar c_5,a_{00})=(0,0).
\end{array}
\end{gather*}
Here $\bar c$'s are arbitrary functions of~$\theta$ with $\bar c_1\bar c_4\frac{\p(\tilde a_2,\dots,\tilde a_r,\tilde a_{00},\tilde b_2)}{\p(a_2,\dots, a_r,a_{00},b_2)}\neq0$ as well as
$\bar X^1(t,\theta)=(\bar c_4\bar T_t)^{-1/\alpha}$ if $\alpha$ is odd or rational in the reduced form with an odd numerator
and $\bar X^1(t,\theta)=\varepsilon|\bar c_4\bar T_t|^{-1/\alpha}$ with $\varepsilon=\pm1$ and $\bar c_4\bar T_t>0$ otherwise.
\end{proposition}

A description of the equivalence group of the class~$\mathcal F_{\mathrm I,00}$ is more complicated and we present
its equivalence groupoid first. In accordance with our standard approach we consider its subclass singled out by the
conditions $a_{00}=b_1=0$.

\begin{proposition}
A point transformation connects the two equations in the class~$\mathcal F_{\mathrm I,00}$ if and only if
its components are of the form
\begin{gather*}
\tilde t=T(t),\quad \tilde x=X^1(t) x,\quad \tilde u=\frac{X^1}{T_t}u-\frac{X^1_{t}}{T_t}x,
\end{gather*}
where $(X^1)^2=c_4T_t$ and the smooth function~$T$ of~$t$ satisfies the equation
\[
\left(\frac{T_{tt}}{T_t}\right)_t-\frac12\left(\frac{T_{tt}}{T_t}\right)^2=2\tilde b_0 T_t^2-2 b_0.
\]
Here $c_4$ is an arbitrary constant and $c_4T_t>0$.
\end{proposition}

The last equation is an autonomous ordinary differential equation on~$T$ which can be integrated in quadra\-tures with standard techniques,
but proceeding this way an explicit form of the general solution thereof can be written only for specific values of parameters.
On the other hand, for any equation in~$\mathcal F_{\mathrm I,00}$ there is an equivalent one to
it in the subclass~$\mathcal F^{b_0=0}_{\mathrm I,00}$ singled out by the condition~$b_0=0$. The corresponding point
transformation is $\tilde t=T(t),\ \tilde x=\sqrt{T_t}x,\ \tilde u=u/\sqrt{T_t}-T_{tt}x/(2\sqrt{(T_t)^3})$,
where a smooth function~$T$ of~$t$ is a solution
of the equation $\left({T_{tt}}/{T_t}\right)_t-\frac12\left({T_{tt}}/{T_t}\right)^2+2 b_0=0$,
for which the general solution can be found explicitly, although a particular solution will suffice for our purposes.
Thus, if $b_0=b^2>0$, then $T(t)=e^{2bt}$ is a particular
solution; if $b_0=-b^2<0$, then $T(t)=\tan(bt)$ is a particular solution, $b>0$ in both cases.

\begin{proposition}
The class~$\mathcal F^{b_0=0}_{\mathrm I,00}$ is normalized in the usual sense. Its usual equivalence group
is constituted by the point transformations of the form
\begin{gather*}
\tilde t=T(t),\quad \tilde x=X^1(t) x,\quad \tilde u=\frac{X^1}{T_t}u-\frac{X^1_{t}}{T_t}x,\quad
\tilde a_2=c_4 a_2,\quad \tilde b_2=c_4^2 b_2,
\end{gather*}
where $X^1(t)=\varepsilon\sqrt{c_4T_t}$ with $\varepsilon=\pm1$,
$T=(c_1t+c_2)/(c_3t+c_0)$ and $c$'s are arbitrary constants, with $\delta=c_1c_0-c_2c_3\neq0$
and $c_0$, $c_1$, $c_2$ and $c_3$ being defined up to a nonzero constant, and $c_4\delta>0$.
\end{proposition}

On the other hand, any admissible transformation of the class~$\mathcal F_{\mathrm I,00}$ can be represented as a composition
of an admissible transformation with a source equation in~$\mathcal F_{\mathrm I,00}$ and a target equation in~$\mathcal F^{b_0=0}_{\mathrm I,00}$, an admissible transformation generated by an equivalence transformation in~$\mathcal F^{b_0=0}_{\mathrm I,00}$ and an admissible transformation back. This way we avoid implicit quadrature expressions arising in the previous approach.
Note that the parameter-function~$T$ is defined as a solution of a third-order ODE parameterized by~$b_0$ and~$\tilde b_0$
and thus should be parameterized by three constants to agree with the Picard--Lindel\"of theorem.
This is indeed the case.

\begin{proposition}\label{Opanasenko:BKdVSt:prop:1}
The class~$\mathcal F_{\mathrm I,00}$ is normalized in the generalized sense. Its effective generalized equivalence
group is constituted by the point transformations of the form
\begin{gather*}
\tilde t=P^2(T(P^1(t))),\quad \tilde x=\sqrt{P^2_{\bar t}P^1_t}X^1(\hat t) x,\\
\tilde u=\frac1{P^2_{\bar t}}\left(\frac{X^1}{T_{\hat t}{P^1_t}}u-\left(\frac{X^1P^1_{t t}}{2T_{\hat t}(P^1_t)^{3/2}}+
\frac{X^1_{\hat t}\sqrt{P^1_t}}{T_{\hat t}}
+\frac{P^2_{\bar t\bar t}X^1\sqrt{P^1_t}}{2P^2_{\bar t}}\right)x\right),\quad
\tilde a_j=c_4 a_j,\quad
\tilde b_2=c_4^2 b_2,\\
\tilde b_0=\frac1{(P^2_{\bar t})^2}\Bigg(
\frac1{(P^1_t)^2}\left(b_0-\left(\frac{P^1_{t t}}{2P^1_t}\right)^2+\frac12\left(\frac{P^1_{t t}}{P^1_t}\right)_t\right)
-\left(\frac{P^2_{\bar t\bar t}}{2P^2_{\bar t}}\right)^2+\frac12\left(\frac{P^2_{\bar t\bar t}}{P^2_{\bar t}}\right)_{\bar t}\Bigg),
\end{gather*}
where $\hat t=P^1(t)$,
$\bar t=T(\hat t)$,
$\tilde t=P^2(\bar t)$,  $X^1(\hat t)=\varepsilon(c_4T_{\hat t})^{1/2}$,
$T=(c_1\hat t+c_2)/(c_3\hat t+c_0)$, with $\delta=c_1c_0-c_2c_3\neq0$,
$c$'s are arbitrary constants,
\[
P^1(t)=\begin{cases}
t&\text{ if }\quad b_0=0,\\
\tan(\sqrt{-b_0}t)&\text{ if }\quad b_0<0,\\
e^{2\sqrt{b_0}t}&\text{ if }\quad b_0>0;
\end{cases}
\]
$P^2(\bar t)$ runs through the set of smooth functions $\{\bar t,\ \frac1{c_5}\ln{|\bar t|},\ \frac1{2c_5}\arctan{\bar t}\}$,
with $c_4\delta>0$, $c_i$, $i=0,1,2,3$, are defined up to a nonzero constant, and $P^2_{\bar t}>0$ and $\varepsilon=\pm1$.
\end{proposition}

The arbitrary element~$\tilde b_0$ of the target equation takes the value of $c_5^2$ if $P^2(y)=\frac1{c_5}\ln{|y|}$,
of $-c_5^2$ if $P^2(y)=\frac1{2c_5}\arctan{y}$ and of~0 otherwise.
The functions $P^2(T(P^1(t)))$ give a three-parameter family of solutions to the nonlinear third-order equation
on~$T$ above parameterized by~$b_0$ and~$\tilde b_0$.

\begin{remark}The point transformations in Proposition~\ref{Opanasenko:BKdVSt:prop:1} form a group by construction, and thus
constitute an effective generalized equivalence group of the class~$\mathcal F_{\mathrm I,00}$. To obtain
the entire generalized equivalence group thereof one allows $c$'s to vary through the set of arbitrary
smooth functions of the arbitrary elements of the class.
\end{remark}

\noindent$\mathbf {II.}$ A class~$\mathcal F_{\mathrm{II}}$ of differential equations of the form
\begin{gather*}
u_t+uu_x=\sum\limits_{j=2}^r a_jx^ju_j+(a_{01}\ln |x|+a_{00})
+x\left(-\frac{a_{01}^2}4\ln^2 |x|
+\left(\frac{a_{01}^2}4-\frac{a_{00}a_{01}}2\right)\ln |x|+b_0\right)
\end{gather*}
is partitioned into the two subclasses~$\mathcal F_{\mathrm{II},0}$ and~$\mathcal F_{\mathrm{II},1}$ that are singled out by conditions $a_{01}=0$ and $a_{01}\neq0$, respectively, and invariant under the admissible transformations of the class~$\mathcal F_{\mathrm{II}}$.

\begin{proposition}
The class~$\mathcal F_{\mathrm{II},0}$ is normalized in the usual sense. Its equivalence group
is constituted by the point transformations of the form
\begin{gather*}
\tilde t=c_1t+c_2,\quad \tilde x=c_4e^{c_3t}x,\quad \tilde u =\frac{c_4e^{c_3t}}{c_1}(u+c_3x),\\
\tilde a_j=\frac {a_j}{c_1},\quad
\tilde {a}_{00}=\frac{a_{00}+2c_3}{c_1},\quad
\tilde b_0=\frac{b_0-c_3^2}{c_1^2},
\end{gather*}
where $c$'s are arbitrary constants with~$c_1c_4\neq0$.
\end{proposition}

The class~$\mathcal F_{\mathrm{II},0}$ is the only owner of a conditional group normalized in the usual sense.

\begin{proposition}
The class~$\mathcal F_{\mathrm{II},1}$ is normalized in the generalized sense.
Its generalized equivalence group~$\bar{G}^\sim_{{\mathrm{II},1}}$ is constituted by the point trans\-forma\-tions of the form
\begin{gather*}
\tilde t=\bar c_1t+\bar c_2,\quad
\tilde x=\bar X^1x,\quad
\tilde u=\frac{\bar X^1}{\bar c_1}\left(u+\frac{\bar c_4a_{01}}2e^{a_{01}t/2}x\right),\quad
\tilde a_j=\frac {a_j}{\bar c_1},\\
\tilde {a}_{01}=\frac {a_{01}}{\bar c_1},\quad
\tilde {a}_{00}=\frac1{\bar c_1}(a_{00}-a_{01} \bar c_3),\quad
\tilde b_0=\frac1{4\bar c_1^2}\left(4b_0-a_{01}^2(\bar c_3^2+\bar c_3)+2a_{00}a_{01}\bar c_3\right),
\end{gather*}
where $\bar X^1:=\exp\left(\bar c_3+\bar c_4\exp(\frac{a_{01}t}2)\right)$,
and $\bar c$'s are smooth functions of the arbitrary ele\-ments $a_{00}$, $a_{01}$, $a_j$ and~$b_0$ with
$\bar c_1\frac{\p(\tilde a_2,\dots,\tilde a_r,\tilde a_{01},\tilde a_{00},\tilde b_0)}{\p(a_2,\dots,a_r,a_{01},a_{00},b_0)}\neq0$.
\end{proposition}

To extract an effective generalized equivalence group from the genera\-lized equivalence group, we set $\bar c_2:=c_2/a_{01}$, $\bar c_3:=-c_3/a_{01}$ and get rid of the dependence of other $\bar c$'s on the arbitrary elements.

\begin{proposition}
An effective generalized equivalence group~$\hat{G}^\sim_{{\mathrm{II},1}}$ of the class~$\mathcal F_{\mathrm{II},1}$ is constituted by the point transformations of the form
\begin{gather*}
\tilde t=c_1t+\frac{c_2}{a_{01}},\quad
\tilde x=X^1(t)x,\quad
\tilde u=\frac{X^1(t)}{c_1}\left(u+\frac{c_4a_{01}}2e^{a_{01}t/2}x\right),\\
\tilde a_j=\frac {a_j}{c_1},\quad
\tilde {a}_{01}=\frac {a_{01}}{c_1},\quad
\tilde {a}_{00}=\frac1{c_1}(a_{00}+ c_3),\quad
\tilde b_0=\frac1{4c_1^2}\left(4b_0+(a_{01}-2a_{00})c_3-c_3^2\right),
\end{gather*}
where $X^1(t):=\exp\big({-}\frac{c_3}{a_{01}}+ c_4\exp(\frac{a_{01}t}2)\big)$
and $c$'s are arbitrary cons\-tants with $c_1\neq0$.
\end{proposition}

The effective generalized equivalence group~$\hat{G}^\sim_{\mathrm{II},1}$ is not a normal subgroup of~$\bar {G}^\sim_{\mathrm{II},1}$,
which is readily seen after writing the time-trans\-forma\-tion out. Therefore,
it is not unique as an effective generalized equivalence group as conjugate subgroups in~$\bar{G}^\sim_{\mathrm{II},1}$ are also
effective generalized equivalence groups. Thus, the existence of a class of differential equations with unique
nontrivial (proper) effective generalized equiva\-lence group is still under question.

\medskip

\noindent$\mathbf{III.}$ A class of differential equations of the form
\begin{gather*}
u_t+uu_x=\sum\limits_{j=2}^r a_je^{\alpha x}u_j+(a_{01}e^{\alpha x}+a_{00})u+b_2e^{2\alpha x}
+b_1e^{\alpha x}+b_0 \quad \text{with } \alpha a_r\neq0
\end{gather*}
admits additional admissible transformations if and only if
\[
b_0=-\frac{a_{00}^2+a_{00}}{2\alpha} \quad\text{ and }\quad b_1=-\frac{a_{00}a_{01}}\alpha.
\]

\begin{proposition}
The class~$\mathcal F_{\mathrm{III}}$ is normalized in the generalized sense. Its generalized equivalence group
is constituted by the point transforma\-tions of the form
\begin{gather*}
\tilde t=\bar T,\quad \tilde x=\bar c_5x-\frac{\bar c_5}{\alpha}\ln|\bar c_4\bar T_t|,\quad \tilde u =
\frac{\bar c_5}{\bar T_t}\left(u-\frac{\bar T_{tt}}{\alpha\bar T_t}\right),\\
\tilde \alpha=\frac{\alpha}{\bar c_5},\quad\tilde a_j=\bar c_4 \bar c_5^ja_j,\quad \tilde a_{01}=\bar c_4a_{01},
\quad \tilde a_{00}=\bar c_3,\quad\tilde b_2=\bar c_4^2\bar c_5b_2,
\end{gather*}
where
the function~$T$ of~$t$ and the arbitrary elements~$\theta$ is defined by
\[
\bar T(t,\theta)=\frac1{\bar c_3}\ln\left|\bar c_3\left(\bar c_1\frac{e^{a_{00}t}-1}{a_{00}}+\bar c_2\right)+1\right|,
\]
and takes the following values at the singular points
\begin{gather*}
\begin{array}{ll}
\bar T(t,\theta)=\bar c_1\dfrac{e^{a_{00}t}-1}{a_{00}}+\bar c_2&\text{ if }\quad \bar c_3=0 \text{ and } a_{00}\neq0,\\[2ex]
\bar T(t,\theta)=\dfrac1{\bar c_3}\ln|\bar c_3(\bar c_1t+\bar c_2)|&\text{ if }\quad a_{00}=0 \text{ and } \bar c_3\neq0,\\[2ex]
\bar T(t,\theta)=\bar c_1{t}+\bar c_2& \text{ if }\quad (a_{00},\bar c_3)=(0,0),
\end{array}
\end{gather*}
$\bar c$'s are smooth functions of~$\theta$
with~$\bar c_1\bar c_4\bar c_5\frac{\p(\bar \alpha,\tilde a_2,\dots,\tilde a_r,\tilde a_{00},\tilde a_{01},\tilde b_2)}{\p(\alpha,a_2,\dots,a_r,a_{00},a_{01},b_2)}\neq0$.
\end{proposition}

To find an effective generalized equivalence group of the class~$\mathcal F_{\mathrm{III}}$ we resort to the following heuristic speculation.
The arbitrary element~$\tilde a_{00}$ may take any real value. Thus, it sufficient to
parameterize $\tilde a_{00}$ to be $a_{00}+c_3$, $c_3\in\mathbb R$. We preserve the number of initial conditions parameterizing~$T$
and guaranteeing the necessary domain for values of~$\tilde a_{00}$. To satisfy another condition of an effective generalized equivalence group
we drop any dependence of remaining $\bar c$'s on the arbitrary elements. In fact, we chose a correct parameterization for them already in the theorem.

\begin{proposition}
An effective generalized equivalence group~$\hat{G}^\sim_{\mathrm{III}}$ of the class~$\mathcal F_{\mathrm{III}}$
is constituted by the point transformations of the form
\begin{gather*}
\tilde t=T,\quad \tilde x=c_5x-\frac{c_5}{\alpha}\ln|c_4 T_t|,\quad \tilde u =
\frac{c_5}{T_t}\left(u-\frac{ T_{tt}}{\alpha T_t}\right),\quad
\tilde \alpha=\frac{\alpha}{ c_5},\\ \tilde a_j=c_4  c_5^ja_j,\quad \tilde a_{01}= c_4a_{01},
\quad \tilde a_{00}=a_{00}+c_3,\quad\tilde b_2= c_4^2 c_5b_2,
\end{gather*}
where the function~$T$ is equal to
\begin{gather*}
T(t)=\frac1{a_{00}+c_3}\ln\left|(a_{00}+c_3)\left(c_1\frac{e^{a_{00}t}-1}{a_{00}}+c_2\right)+1\right|,
\end{gather*}
and takes the following values at the singular points
\begin{gather*}
\begin{array}{ll}
T(t)=c_1\frac{e^{a_{00}t}-1}{a_{00}}+c_2&\text{ if }\quad  c_3=- a_{00}\neq0,\\[2ex]
T(t)=\frac1{c_3}\ln| c_3( c_1t+ c_2)|&\text{ if }\quad a_{00}=0 \text{ and }  c_3\neq0,\\[2ex]
T(t)=c_1{t}+ c_2 & \text{ if }\quad(a_{00},c_3)=(0,0),
\end{array}
\end{gather*}
and $c$'s are arbitrary constants with~$ c_1 c_4 c_5\neq0$.
\end{proposition}

Guided by the same logic as for the class~$\mathcal F_{\mathrm{II},1}$, we can show nonuniqueness of effective generalized
equivalence groups for~$\mathcal F_{\mathrm{III}}$ as well.

\medskip

\noindent$\mathbf{IV.}$ Finally we discuss the last subclass~$\mathcal F_{\mathrm {IV}}$ of~$\mathcal F$ admitting additional admissible transformations.
It consists of the equations
\[
u_t+uu_x=\sum_{j=2}^r a_ju_j+a_0u+b_1x+b_0.
\]
Since the arbitrary elements $a_j$ are scaled under the action of the equivalence group of the class, it is
reasonable to single out two subclasses of the class under question: $\mathcal F_{\mathrm {IV},0}$ with $a_j=0$ for all~$j=2,\dots,r-1$,
and complementary to it the subclass~$\mathcal F_{\mathrm {IV},1}$ with at least one $a_j$ nonzero.

\begin{proposition}The class~$\mathcal F_{\mathrm{IV},1}$ is normalized in the generalized sense. Its generalized equivalence group
is constituted by the point transforma\-tions of the form
\begin{gather*}
\tilde t=\bar T^1t+\bar T^0,\quad \tilde x=\bar X^1x+\bar X^0,\quad \tilde u =\frac{\bar X^1}{\bar T^1}u+\frac{\bar X^0_t}{\bar T^1},\\
\tilde a_j=\frac{(\bar X^1)^r}{\bar T^1}a_j,\quad
a_0=\frac{a_0}{\bar T^1},\quad \tilde b_{1}=\frac{b_1}{(\bar T^1)^2},
\quad\tilde b_{0}=\frac1{(\bar T^1)^2}\left(\bar X^1b_0+\bar c_3\right),
\end{gather*}
where
\[
\bar X^0(t,\theta)=
\begin{cases}
\bar c_1e^{\lambda_1t}+\bar c_2e^{\lambda_2t}+\bar c_3 & \text{ if } \quad\lambda_1\neq0, \quad D>0\\
\bar c_1t+\bar c_2e^{\lambda_2t}+\bar c_3 & \text{ if } \quad\lambda_1=0, \quad D>0,\\
\bar c_1e^{b_1t/2}+\bar c_2te^{b_1t/2}+\bar c_3 & \text{ if }\quad b_1\neq0,\quad D=0\\
\bar c_1t^2+\bar c_2t+\bar c_3 & \text{ if } \quad b_1=0, \quad D=0,\\
e^{b_1t/2}\big(\bar c_1\sin(\sqrt{-D}t)+\bar c_2\cos(\sqrt{-D}t)\big)+\bar c_3 & \text{ if }\quad D<0,
\end{cases}
\]
where $D=b_1^2+4a_0$ and $\lambda_{1,2}=(b_1\pm\sqrt{D})/2$ with $|\lambda_1|<|\lambda_2|$,
$\bar X^1$, $\bar T^0$, $\bar T^1$ and $\bar c $'s run through the set smooth functions of
the arbitrary elements~$\theta=(a_j,a_0,b_1,b_0)$ with~$\bar X^1\bar T^1\frac{\p(\tilde a_2,\dots,\tilde a_r,\tilde a_0,\tilde b_1,\tilde b_0)}{\p(a_2,\dots,a_r,a_0,b_1,b_0)}\neq0$.
\end{proposition}

Here the function~$X^0(t)$ is a solution of the ordinary differential equation $X^0_{ttt}-b_1X^0_{tt}-a_0X^0_t=0$ and thus
it smoothly depends on the parameters~$b_1$, $a_0$ and all the initial conditions.

The equivalence groupoid of the class~$\mathcal F_{\mathrm {IV},0}$ depends essentially on the order~$r$ of equations therein.
So we consider both the cases separately. First assume that $r>2$ and denote the class of such
equations~$\mathcal F^{r>2}_{\mathrm {IV},0}$. This class admits additional admissible transformations if and only if $b_1=a_0^2(r-1)/(r-2)^2$, so we reduce a tuple of the arbitrary elements thereof by the element~$b_1$.

\begin{proposition}
The class~$\mathcal F^{r>2}_{\mathrm {IV},0}$ is normalized in the generalized sense. Its generalized equivalence group
is constituted by the point transformations of the form
\begin{gather*}
\tilde t=\bar T,\quad \tilde x=\bar X^1x+\bar X^0,\quad \tilde u =\frac{\bar X^1}{\bar T_t}u+\frac{\bar X^1_t}{\bar T_t}x+\frac{\bar X^0_t}{\bar T_t},\quad
\tilde a_r=\frac{(\bar X^1)^r}{\bar T_t}a_r,
\quad \tilde a_0=\bar c_3, \quad \tilde b_0=\bar c_5,
\end{gather*}
where a pair of smooth functions~$(\bar T,\bar X^0)$ of~$t$ and the arbitrary elements~$\theta$ equal to
\begin{gather*}
(\bar c_1t+\bar c_2,\ \bar c_7t^2+\bar c_6t+\bar c_5) \quad \text { if }\quad a_0=0\text{ and }\bar c_3=0,\\[1ex]
\bigg(\frac1{\bar c_3}\ln|\bar c_3(\bar c_1t+\bar c_2)|,\
\-\frac{\bar c_5r^2}{\bar c_3^2(r-1)}+\frac{\bar c_6t+\bar c_7}{\left|t+{\bar c_2}/(\bar c_1\bar c_3)\right|^{1/r}}\\
\qquad\qquad\qquad\qquad\quad{}-\frac{\bar c_3^2b_0\bar X^1}{2}\left(t+\frac{\bar c_2}{\bar c_1\bar c_3}\right)^2\bigg)
\  \text { if } a_0=0\text{ and }\bar c_3\neq0,\\[1ex]
\Big( \frac{r-2}{\bar c_3 r}\ln\left|\frac{1}{\bar c_1}e^{\frac{a_0rt}{r-2}}+\frac{\bar c_2}{\bar c_1}\right|,\
\-\frac{\bar c_5(r-2)^2}{\bar c_3^2(r-1)} +\frac{\left(\bar c_6e^{\frac{a_0rt}{r-2}}+\bar c_7\right)}
{\left|{\bar c_2}/{\bar c_1}+e^{\frac{a_0rt}{r-2}}\right|^{1/r}}+\frac{(r-2)^2b_0}{(r-1)a_0^2}\bar X^1\Big)
\ \ \text { if }\ \ a_0\bar c_3\neq0,\\[1ex]
\left(\bar c_1e^{\frac{a_{0}rt}{r-2}}+\bar c_2, \ \frac{\bar c_5\bar c_1^2}2e^{\frac{2a_0rt}{r-2}}+\bar c_6e^{\frac{a_0rt}{r-2}}
+\bar c_7-\frac{(r-2)^2b_0}{(r-1)a_0^2}\bar X^1\right) 
\quad \text{ if }\quad a_0\neq0\text{  and  }\bar c_3=0,
\end{gather*}
where $\bar c$'s are arbitrary smooth functions of~$\theta$ with $\bar c_4\bar T_t\frac{\p(\tilde a_r,\tilde a_0,\tilde b_0)}{\p(a_r,a_0,b_0)}\neq0$ as well as
$\bar X^1(t,\theta)=\varepsilon(\bar c_4\bar T_t)^{1/r}$ with $\varepsilon=\pm1$ and $\bar c_4\bar T_t>0$
if $r$ is even and $\varepsilon=1$ otherwise.
\end{proposition}

The second function in the second pair in the second set gives a~ge\-ne\-ral solution of the linear inhomogeneous equation on~$X^0(t)$,
\[
\bar c_5=b_0\frac{X^1}{T_t{}^2}+\frac1{T_t}\left(\frac{X^0_t}{T_t}\right)_t-\bar c_3\frac{X^0_t}{T_t}-\frac{\bar c_3^2(r-1)}{(r-2)^2}X^0,
\]
parameterized by the first function~$T$ in the set and the corresponding $X^1(t)$. Any particular solution of this equation seems impossible
to find with standard techniques. Here instead, we used a method used for the class~$\mathcal F_{\mathrm {I},00}$ with gauging the arbitrary
elements $a_0$ and~$b_0$ to~0 first and composing equivalence transformations thereafter.

Due to the above condition on the arbitrary elements~$b_1$ and~$a_0$, the class~$\mathcal F^{r=2}_{\mathrm{IV},0}$ admits
additional admissible transformations if and only if $a_0=0$. Abusing notations we denote the subclass singled out by this
condition again by~$\mathcal F^{r=2}_{\mathrm{IV},0}$.

\begin{proposition}
The point transformation of the form
\[
\tilde t=T(t),\quad \tilde x=X^1(t)x+X^0(t),\quad \tilde u=\frac{X^1}{T_t}u+\frac{X^1_t}{T_t}x+\frac{X^0_t}{T_t}
\]
connects the source and target equations in the class~$\mathcal F^{r=2}_{\mathrm{IV},0}$ if and only if
$(X^1)^2/T_t=\operatorname{const}\ne0$, the parameter function~$T$ runs through the solution set of the system
\[
\left(\frac{T_{tt}}{T_t}\right)_t-\frac12\left(\frac{T_{tt}}{T_t}\right)^2=
2\tilde b_1 T_t^2-2 b_1,
\]
and the parameter function $X^0$ of~$t$ satisfies the equation
\[
\frac1{T_t}\left(\frac{X^0_t}{T_t}\right)_t-\tilde b_1X^0=\tilde b_0-b_0\frac{X^1}{T_t{}^2}.
\]
\end{proposition}

The last equation is linear inhomogeneous with respect to $X^0(T)$ for a given $T(t)$, while the differential
equation on~$T$ is integrated in quadratures as an autonomous equation on $\ln|T_t|$ with standard techniques.
Nonetheless, using the similar trick as was used for the class~$\mathcal F_{\mathrm I,00}$, one can do better.
More precisely, we gauge the arbitrary
elements $b_0$ and~$b_1$ to zeros by the point transformation of the form
\[
\tilde t=T(t),\quad \tilde x=\sqrt{T_t}x+X^0(t),\quad \tilde u=\frac u{\sqrt{T_t}}+\frac{T_{tt}x}{2(T_t)^{3/2}}+\frac{X^0_t}{T_t},
\]
where
\begin{gather*}
\begin{array}{ll}
\big(T,X^0\big)=\big(e^{2\sqrt{b_1}t},\ 4b_0(2\sqrt{b_1})^{3/2}e^{\sqrt{b_1}t}\big)&  \text{ if }\quad b_1>0;\\[1ex]
\big(T,X^0\big)=\left(\tan(\sqrt{-b_1}t),\ \dfrac{-b_0(-b_1)^{3/4}}{\cos{\sqrt{-b_1}t}}\right)& \text{ if }\quad b_1<0,
\end{array}
\end{gather*}
obtaining the subclass~$\mathcal F^{r=2}_{\mathrm{IV},00}$ of~$\mathcal F^{r=2}_{\mathrm{IV},0}$.
Thereafter we present the equivalence groupoid of~$\mathcal F^{r=2}_{\mathrm{IV},0}$ by composing an equivalence transformation within
the subclass~$\mathcal F^{r=2}_{\mathrm{IV},00}$ with point transformations mapping equations in the superclass
to equations in the subclass and back.

\begin{proposition}
The class~$\mathcal F^{r=2}_{\mathrm {IV},00}$ is normalized in the usual sense.
Its usual equivalence group is constituted by point transformations of the form
\begin{gather*}
\tilde t=T(t),\quad \tilde x=X^1(t)x+X^0,\quad \tilde u =\frac{X^1}{T_t}u+\frac{X^1_t}{T_t}x,\quad
\tilde a_2=c_4a_2,
\end{gather*}
where $X^1(t)=\varepsilon(c_4\delta)^{1/2}/(c_3t+c_0)$,
$T=(c_1t+c_2)/(c_3t+c_0)$,
$X^0$ and $c$'s are arbitrary constants with $\delta=c_1c_0-c_2c_3\neq0$, $c_0$, $c_1$, $c_2$ and $c_3$ being defined up to a nonzero constant, $c_4\delta>0$ and $\varepsilon=\pm1$.
\end{proposition}

The point transformation~$\mathcal T_{\tilde T,\tilde X^0}$ which maps an equation in~$\mathcal F^{r=2}_{\mathrm {IV},00}$ to an equation
in~$\mathcal F^{r=2}_{\mathrm {IV},0}$ is of the same form as above,
\[
t=\tilde T(\tilde t),\quad  x=\sqrt{|\tilde T_{\tilde t}|}\tilde x+\tilde X^0(\tilde t),\quad  u=
\frac{\tilde u}{\sqrt{|\tilde T_{\tilde t}|}}+\frac{\tilde T_{\tilde t\tilde t}\tilde x}{2|\tilde T_{\tilde t}|^{3/2}}+
\frac{\tilde X^0_{\tilde t}}{\tilde T_{\tilde t}},
\]
where $\tilde T(T(t))=t$ and $\tilde X^0(\tilde t)=-(X^0/T_t)(\tilde T(\tilde t))$, that is,
\begin{gather*}
\begin{array}{ll}
\big(\tilde T,\tilde X^0\big)=\left(\dfrac{\ln|\tilde t|}{2\sqrt{\tilde b_1}},\
\dfrac{\tilde b_0}{\tilde b_1}\right)  & \text{if} \quad\tilde b_1>0;\\[1ex]
\big(\tilde T,\tilde X^0\big)=\left(\dfrac{\arctan(\tilde t)}{\sqrt{-\tilde b_1}},\
\dfrac{\tilde b_0}{\tilde b_1}\right) &\text{if}\quad \tilde b_1<0.
\end{array}
\end{gather*}

\begin{proposition}\label{Opanasenko:StBKdV:prop26}
The class~$\mathcal F^{r=2}_{\mathrm IV,0}$ is normalized in the generalized sense. Its effective generalized equivalence
group is constituted by the point transformations of the form
\begin{gather*}
\tilde t=P^2(T(P^1(t))),\quad
\tilde x=\sqrt{P^2_{\bar t}P^1_t}X^1(\hat t) x+\sqrt{P^2_{\bar t}}(X^1R^1+X^0)+R^2,\\
\tilde u=\frac{X^1u}{\sqrt{P^2_{\bar t}P^1_t}T_{\hat t}}+
\left(\frac{X^1P^1_{t t}}{2T_{\hat t}\sqrt{P^2_{\bar t}(P^1_t)^3}}+ \frac{X^1_{\hat t}\sqrt{P^1_t}}{\sqrt{P^2_{\bar t}}T_{\hat t}}
+\frac{P^2_{\bar t\bar t}X^1}{2(P^2_{\bar t})^2\sqrt{P^1_t}}\right)x\\
\hphantom{\tilde u=}+\frac{X^1R^1_t}{T_{\hat t}\sqrt{P^2_{\bar t}}P^1_t}+\frac{X^1_{\hat t}R^1}{T_{\hat t}}+
\frac{P^2_{\bar t\bar t}}{2(P^2_{\bar t})^{3/2}}\left(X^1R^1+X^0\right)+\frac{R^2_{\bar t}}{P^2_{\bar t}},
\quad\tilde a_2=c_4 a_2,\\
\tilde b_1=\frac1{(P^2_{\bar t})^2}\Bigg(
\frac1{(P^1_t)^2}\left(b_1+\frac12\left(\frac{P^1_{tt}}{P^1_{t}}\right)_t-\frac14\left(\frac{P^1_{tt}}{P^1_{t}}\right)^2\right)
+\frac12\left(\frac{P^2_{\bar t\bar t}}{P^2_{\bar t}}\right)_{\bar t}-
\frac14\left(\frac{P^2_{\bar t\bar t}}{P^2_{\bar t}}\right)^2\Bigg),\\
\tilde b_0=\frac1{(P^2_{\bar t})^{3/2}}\left(\frac{b_0}{(P^1_t)^{3/2}}+
\frac1{P^1_{t}}\left(\frac{R^1_{t}}{P^1_{t}}\right)_{t}\right)
+\frac1{P^2_{\bar t}}\left(\frac{R^2_{\bar t}}{P^2_{\bar t}}\right)_{\bar t}-\tilde b_1R^2,
\end{gather*}
where $\hat t=P^1(t)$,
$\bar t=T(\hat t)$,
$\tilde t=P^2(\bar t)$,
$X^1(\hat t)=\varepsilon(c_4T_{\hat t})^{1/2}$,
$T=(c_1\hat t+c_2)/(c_3\hat t+c_0)$ with $\delta=c_1c_0-c_2c_3\neq0$;
\[
(P^1(t),R^1(t))=
\begin{cases}
(t,\ -b_0t^2/2) &\text{ if } b_1=0, \\
(\tan(\sqrt{-b_1}t),\ -b_0(-b_1)^{3/4}/\cos(\sqrt{-b_1}t)) &\text{ if } b_1<0,\\
(e^{2\sqrt{b_1}t},\ 4b_0(2\sqrt{b_1})^{-3/2}e^{\sqrt{b_1}t}) &\text{ if } b_1>0;
\end{cases}
\]
$X^0$ and $c$'s are arbitrary constants and
the pair of smooth functions $(P^2(\bar t),R^2(\bar t))$ runs through the set
\[
\left\{\left(\bar t,\ \frac{c_6\bar t^2}2\right),\ \left(\frac{\ln{|\bar t|}}{2c_5},\ \frac{c_6}{c_5^2}\right),\
\left(\frac{\arctan{\bar t}}{c_5},\ -\frac{c_6}{c_5^2}\right)\right\},
\]
with $c_i$, $i=0,\dots,3$, being defined up to a nonzero constant, $c_4\delta>0$, $P^2_{\bar t}>0$ and $\varepsilon=\pm1$.
\end{proposition}

In the notation of Proposition~\ref{Opanasenko:StBKdV:prop26}, the point transformation~$\mathcal T_{P^2,R^2}$
maps an equation in~$\mathcal F^{r=2}_{\mathrm {IV},00}$
to an equation in~$\mathcal F^{r=2}_{\mathrm {IV},0}$ with arbitrary-element tuples~$(b_0,b_1)$ equal to $(c_6,0)$, $(c_6,c_5^2)$ and $(c_6,-c_5^2)$, respectively.

\section*{Acknowledgments}
The author is grateful to Roman O.~Popovych for useful discussions and interesting comments,
and acknowledges the support of the Natural Sciences and Engineering Research Council of Canada.

\footnotesize\setlength{\itemsep}{0.3ex} \label{Opanasenko:LastPage}

\newpage


\frenchspacing

\begin{thebibliography}{10}

\bibitem{Opanasenko:Arnold1992}
Arnol'd V.I., Ordinary differential equations, Springer-Verlag, Berlin, 1992.

\bibitem{Opanasenko:kuru16a}
Kurujyibwami C., Basarab-Horwath P., Popovych R.O., Algebraic method for
  group classification of~(1+1)-dimensional linear {S}chr\"odinger equations,
  \emph{Acta Appl. Math.} \textbf{157} (2018), 171--203, arXiv:1607.04118.

\bibitem{Opanasenko:mele94Ay}
Meleshko S.V., Group classification of equations of two-dimensional gas
  motions, {\it J.~Appl. Math. Mech.}  \textbf{58} (1994), 629--635.

\bibitem{Opanasenko:mele96Ay}
Meleshko S.V., {Generalization of the equivalence transformations}, \emph{J.~Nonlinear Math. Phys.} \textbf{3} (1996), 170--174.

\bibitem{Opanasenko:NikPop}
Nikitin A.G., Popovych R.O.,
Group classification of nonlinear Schr\"odinger equations,
\emph{Ukrainian Math. J.} \textbf{53} (2001), 1255--1265, arXiv:math-ph/0301009.

\bibitem{Opanasenko:Opanasenko17}
Opanasenko S., Bihlo A., Popovych R.O.,
Group analysis of general {B}urgers--{K}orteweg--de {V}ries equations,
\emph{J.~Math. Phys.} \textbf{58} (2017), 081511, 37~pp., arXiv:1703.06932.

\bibitem{Opanasenko:Opanasenko19}
Opanasenko S., Boyko V., Popovych R.O.,
Enhanced group classifica\-tion of reaction-diffusion equations with gradient-dependent diffusion,
 arXiv:1804.08776.

\bibitem{Opanasenko:popo06Ay}
Popovych R.O., Classification of admissible transformations of differential
  equations, in \emph{Collection of Works of Institute of Mathematics},
  vol.~3, Institute of Mathematics, Kyiv, 2006, 239--254.

\bibitem{Opanasenko:popo10Ay}
Popovych R.O., Kunzinger M., Eshraghi H., Admissible transformations and
  normalized classes of nonlinear Schr\"odinger equations, \emph{Acta Appl.
  Math.} \textbf{109} (2010), 315--359, arXiv:math-ph/0611061.

\bibitem{Opanasenko:vane07Ay}
Vaneeva O.O., Johnpillai A.G., Popovych R.O., Sophocleous C., Enhanced group
  analysis and conservation laws of variable coefficient reaction-diffusion
  equations with power nonlinearities, \emph{J.~Math. Anal. Appl.} \textbf{330}
  (2007), 1363--1386, arXiv:math-ph/0605081.

\end{thebibliography}
\end{document}